# Comparative Analysis for Quick Sort Algorithm under four Different Modes of Execution

Mohammad Fasha, January 2014

**Abstract**

This work presents a comparison for the performance of sequential sorting algorithms under four different modes of execution, the sequential processing mode, a conventional multi-threading implementation, multi-threading with OpenMP Library and finally parallel processing on a super computer. Quick Sort algorithm was selected to run the experiments performed by this effort and the algorithm was run using different arrays sizes and different number of processors. The results and findings were analyzed uncovering limitations as well as enhancement potentials of sequential sorting algorithms using parallelism.

**Introduction**

Sequential sorting algorithms are known for their efficiency and simplicity when sorting arrays and datasets of different sizes. The time complexity of these algorithms spans from O (n Log n) to O ($n^2$). For larger datasets, these algorithms disclose performance degrades due to the relatively large number of computations and iterations that need to be performed sequentially.

The work presented in this paper examined performance capabilities and limitations of quick sort sequential algorithm when sorting integer arrays of different sizes. Mainly, the presented work can be divided into two different sections, the first section relates to the experiments that were run on a single multi-core machine, while the second part concerns the experiments that were run on (IMAN1), a high end super computer owned and operated by the government of Jordan.

As for the single machine experiments, three different execution scenarios were experimented, the first was running a conventional sequential version of quick sort algorithm, in the second run, quick sort algorithm was tested after introducing conventional multi-threading capabilities while in the third and last experiment that was run on the single machine, an OpenMP version of the multi-threaded algorithm was experimented disclosing significant enhancements witnessed as will be discussed in the next sections. Results and findings for the single machine experiments were collected and analyze.

In the second section of this work, quick sort algorithm was tested on (IMAN1) super computer. Results and findings were also captured and analyzed and compared to the results and findings of the first section experiments. In the next sections, a detailed discussion about the experiments is provided and an analysis of the results is presented.

**Quick Sort**

Quick sort is a divide-and-conquer sorting algorithm that was developed in the 1960s [1]. This algorithm works by partitioning a given dataset into two parts based on a selected pivot, the low part elements contain values less than the selected pivot while the high part contain values that are higher than the pivot. The algorithm is recursively called with the same concept over every two partitions until no further division is possible, after which, the array will be sorted. Quick sort is known to perform $O(n \log n)$ in average and $O(n^2)$ in worst case scenarios [2]. Figure 1 and 2 below present the pseudo code and the general flow of the algorithm.

*int function Partition (Array A, int Lb, int Ub);*
    *begin*
        *select a pivot from A[Lb]…A[Ub];*
        *reorder A[Lb]…A[Ub] such that:*
        *all values to the left of the pivot are ≤ pivot*
        *all values to the right of the pivot are ≥ pivot*
        *return pivot position;*
    *end;*

*Procedure QuickSort (Array A, int Lb, int Ub);*
    *begin*
        *if Lb < Ub then*
        *M = Partition (A, Lb, Ub);*

        *QuickSort (A, Lb, M – 1);*
        *QuickSort (A, M + 1, Ub);*
    *end;*

*Figure -1- Pseudo Code of Quick Sort Algorithm disclosing its simple implementation and its recursive nature [2]*

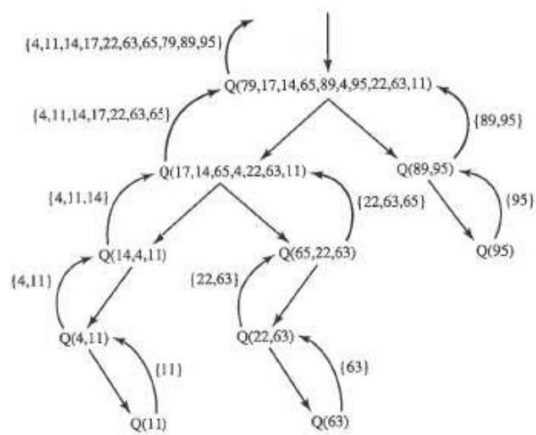

*Figure -2- General processing flow of the Quick Sort Algorithm showing how arrays are continuously and recursively split into two parts based on their value according to a selected pivot [3]*

**Experimentations and Results**

**Phase One, Running experiments on a single machine**

In the first phase of this work, quick sort algorithm was experimented on a single machine with the following specifications:

- Intel Core i 7 @ 2.2 GHz dual (quad cores).
- 6 GB Ram.
- Windows 7 OS 64 bit version.
- GNU C++ version 4.7 using Code Blocks IDE.

Three different types of scenarios were run on this single machine, these are, 1-The conventional sequential quick sort, 2-The conventional multi-threaded quicksort and finally 3-An OpenMp version of the multi-threaded scenario.

**Conventional Sequential Mode:**

In this scenario, conventional quick sort algorithm was experimented on the aforementioned environment using arrays of different sizes [1 million, 5 million, 10 million, 50 million, 100 million and 400.000.000 integers]. Also, those arrays were tested after being initialized with an already sorted data (ascending or descending) as well as randomly generated ones. Pivot in the middle scheme was used to run the algorithm. As presented in Figure-3 below, the algorithm performed better over the already sorted arrays while in the random generated ones, the time consumption was noticeably higher.

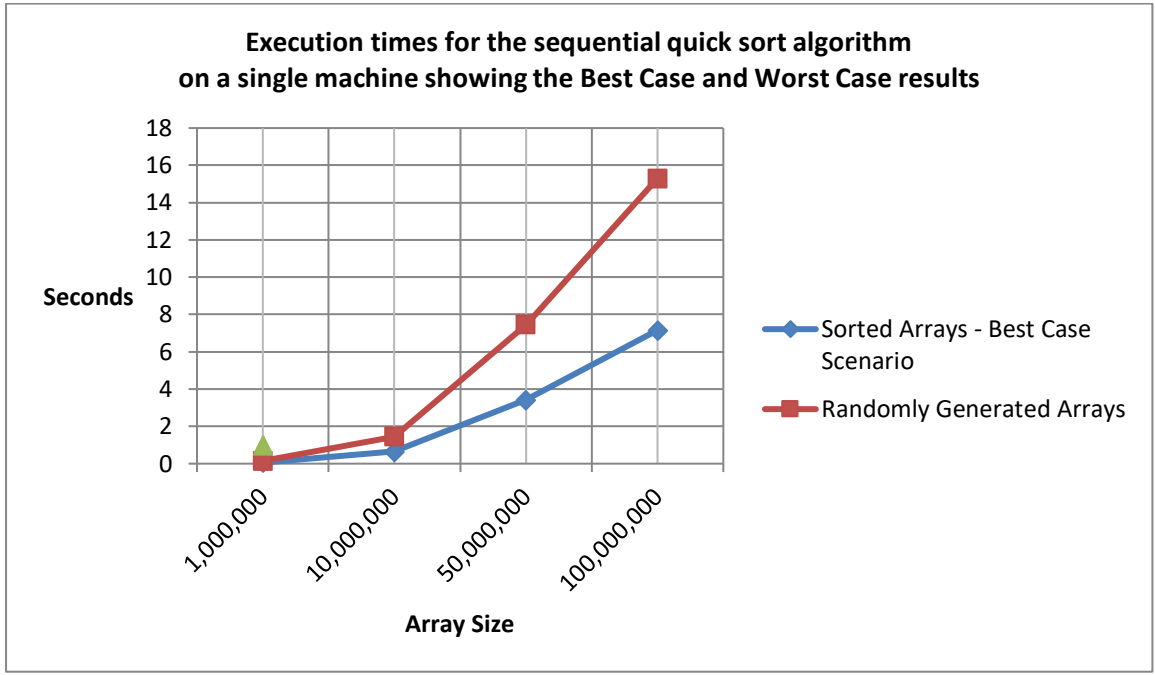

*Figure 3 - Execution times in seconds when running a conventional version of the sequential quick sort algorithm on a single machine using integer arrays of different sizes, Table 1 in Appendix [A] presents the actual execution times for this graph.*

A noticeable fact about the previous scenario was that the number of iterations under the already sorted arrays was higher than their counter parts in the randomly generated ones, while the execution time under the already sorted arrays was noticeably less than the execution times for the randomly generated arrays. This can be explained after examining Table 2, Table 3 and Table 4 in Appendix [A] which clarifies that the number of swaps and the number of recursion calls in the randomly generated arrays scenarios were much higher than their counter parts in the already sorted arrays. These additional execution steps caused the execution time to be higher in the randomly generated arrays scenarios than the execution times in the already sorted arrays scenarios.

**Conventional Multi-Threading Mode**

In the second run of the algorithm, conventional multi-threading was experimented.

Quick sort algorithm has inherent characteristics that make it suitable for parallelization. This algorithm is of divide-and-conquer nature, therefore, its initial dataset can be divided into smaller data sets that are explicitly dispatched to remote processors for parallel execution, once the execution is finished at a given distributed processor, the sorted sub-arrays are sent back to a central processing node which merges the results and generates the original dataset with all its items being sorted.

The same environment was used to run MPI based sort. The used version of Windows GNU C++ compiler has built-in support for OpenMP libraries. To run this scenario, the initial array was initialized as an already-sorted array which was divided into (8) parts according to the number of the available CPUs and Cores (2 x 4) on the test environment.

The program flow was as following:

1. Partitioning the data to create a number of subsets compatible with the number of the available processors, (8) sub-arrays in our case.
2. Create a number of threads according to the number of the available processors, (8) threads in our case.
3. Assign a sub-array to each of the (8) threads.
4. Each thread starts sorting its designated subset sequentially and all the threads perform their work in parallel.

Once all the (8) threads have finished execution, sub-arrays merging would be needed.

The multi-threading scenario demonstrated partial (minor) enhancement over the previous sequential processing scenario. This can be explained because underneath, Windows 7 was running the (8) threads of execution over an undetermined number of CPUs and Cores, which was most probably less than the available full processing power of the machine. The previous justification was supported after the examination of CPU utilization using Windows OS CPU utilization probe. Figure 1 in Appendix [B] presents a snapshot of the CPU utilization screen which was captured during the execution of the algorithm. That snapshot clearly demonstrates that the CPUs utilization was around (12%) of the total power of the machine when running the algorithm using the conventional multi-threaded scenario.

**OpenMP Multi-Threading Mode**

The third scenario that was run on the single machine was similar to the previous conventional multi-threaded scenario, only this time, each thread of execution was explicitly assigned a dedicated processor using the OpenMP C library.

In this scenario, significant performance enhancement was witnessed as presented in Figure-4 below. The enhancement in performance can be related to the same reasons that were clarified in the previous section. In this scenario, CPU utilization on the test machine reached its full processing powers (100%) during the execution of the program. A snapshot of the CPU utilization under this situation is presented in Figure-2 in Appendix [B].

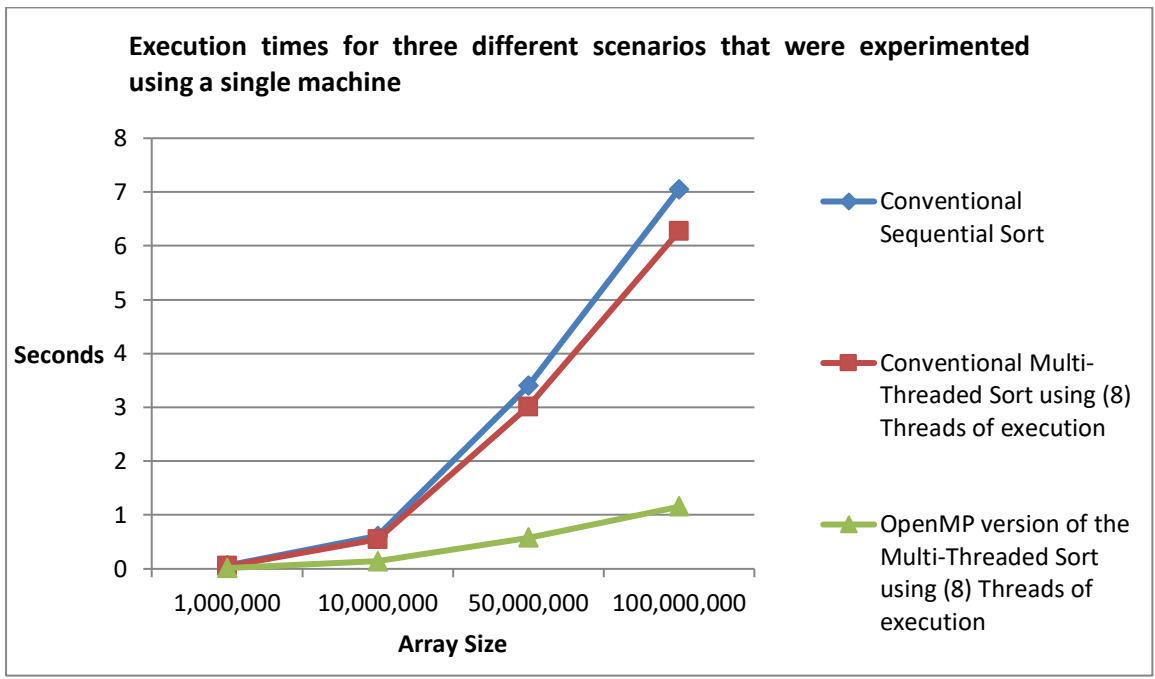

*Figure 4 - Execution times in seconds for the three scenarios that were run on a single machine using different array sizes, 1-The conventional sequential algorithm 2-Conventional multi-threading using (8) threads of execution 3-OpenMP version of the multi-threaded algorithm using (8) threads of execution. The actual data for this graph is presented in Table 5 in Appendix [A].*

The code adjustment for enabling OpenMP multi-threading is presented in Figure-4 below. This code snippet demonstrates how the algorithm assigned a subset of the main array to a dedicated thread of execution.

```
int tid;
  /* Fork a team of threads giving them their own copies of variables */
    #pragma omp parallel private(tid)
    {
       QuickSorter innerQuickSorter;
       /* Obtain thread number */
       tid = omp_get_thread_num();
          innerQuickSorter.RunSortPivotMiddle(MainArray,SubArraysInfo[tid].lower_bound, SubArraysInfo[tid].upper_bound);
         //printf("Thread number %d finished sorting elements %d to %d\n", tid, SubArraysInfo[tid].lower_bound, SubArraysInfo[tid].upper_bound);
     } /* All threads join master thread and disband */
```

*Figure 5 – The simple C OpenMP code snippet that was used to enforce the usage of all the available processors on the test machine.*

**Super Computer Experiments**

In the second phase of this work, An MPI version of the algorithm was experimented on IMAN1 super computer. IMAN1 is a supercomputer located in the Kingdom of Jordan that was locally built using (2260) PlayStation3 machines. This parallel computing machine produces a computing power of (25) TeraFLOPS. The design theme of IMAN1 was to establish a powerful parallel processing hub using reasonable costs.

In this super computer, each play station machine has a single PPE (Power Processing Element) processor with the following specifications:

- 64-bit PowerPC
- SMP (2 threads)
- 3.2 GHz
- 256MB XDR DRAM

It also has (8) SPEs (synergistic Processing Element) each with the following specifications:

- 256 KB SRAM
- 3.2 GHz with VMX vector unit
- 128-bit Vector Registers

The computer was built using YellowDog linux OS, it has an open source MPI-2 implementation as well as the GNU compilers collection. The experiments that were run on IMAN1 didn't use the full power of this supercomputing, but rather a fraction of its processing powers was utilized. The experiments were run using (128) PlayStation nodes. Moreover, the powerful (8) SPEs of each PlayStation3 box weren't used; only the single PPEs in each node were utilized.

For our experiments, one of the PlayStation machines was selected to perform the role of the master node that will do the following tasks once the program is run:

1. Initialize a random array of a given size.
2. Rearrange the array according to the number of available processors. The difference between this scenario and the last scenario of the previous section was in the array organization process. In this scenario, the master array was split into a number of sub arrays according to the number of the available processors, taking into consideration that the array division was organized according to a number of pivots equal to the number of the available processors. The arrangement allowed the dispatching of pre-organized sub-arrays that can be remotely sorted and collected back to perform a simple concatenation rather than full sub-arrays merge.
3. Dispatch the sub-arrays to remote processors.
4. Collected back the sorted sub-arrays and perform simple concatenation.

The general flow of the program was as following:

1. At the master node, initialize a random integer array with different number of integers in each scenario [1000000, 5000000, 10000000, 20000000, 50000000…etc.].
2. Divide the master array according to the aforementioned paragraph.
3. Dispatch the generated sub-arrays to the remote processors.
4. Once a remote processor receives its designated sub array, it will start sorting that segment using the conventional sequential quick sort algorithm.
5. Once a remote node concludes sequentially sorting its designated sub-array, it shall send its local execution time as well as the resulted sorted array back to the master node.
6. All the sorted sub arrays are received back by the master node, and the time results are computed and displayed.

Figure-6 below presents the initial results when running the quick sort algorithm on IMAN1. As presented in the diagram, execution time increases with the increase in the number processors, a rather abnormal behavior opposite to what it is expected. Nevertheless, this behavior

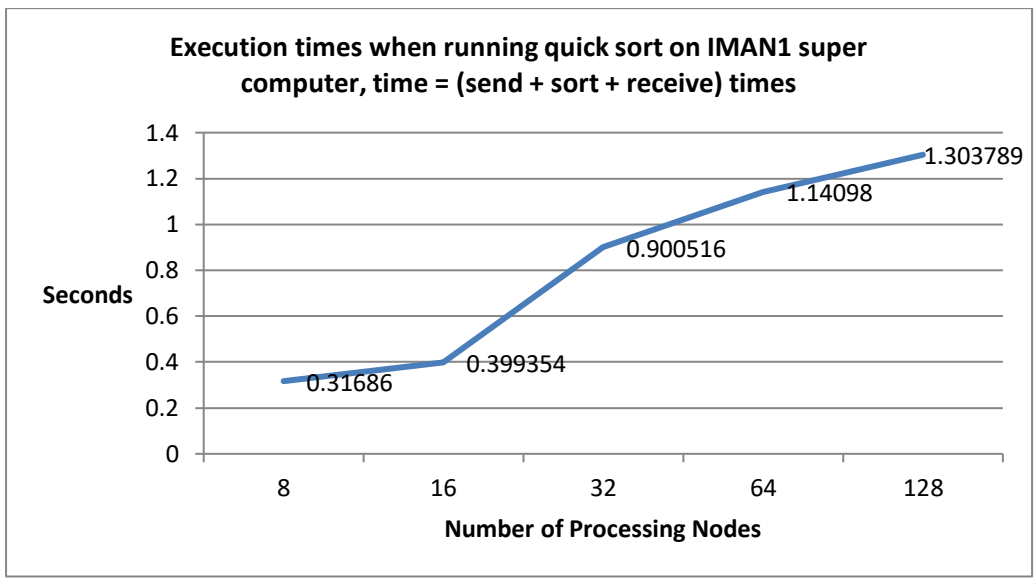

*Figure 6 – Execution time for the quick sort algorithm on IMAN1 super computer. The experiments were run using an initial array size of 5000000 elements while using different number of processors ranging from 8 to 128 nodes.*

can be explained once we clarify that the resulted execution time in our experiments included the communication time that was required for sending and receiving back the results. It is also important to clarify that during the experiments, numerous challenges were met because of the performance limitations of the master node. For our experiments, it was decided to use one of the available PlayStations as the master node. The master node duties that were previously discussed were assigned to this low end (256MB RAM) machine. This memory limitation increased the number of swaps towards the hard disks and the virtual memory space which was visible when we examined the master node designated LEDS and the memory listings during the execution. The numerous duties of the master node consumed more than (220 Seconds) in some situations causing a noticeable degrade in performance. The performance degrade was apparent especially when using main arrays with sizes > 5 million integers. Table-6 in Appendix [B] presents some of the results that demonstrate the time consumed during the array preparation and sending process.

To determine the actual execution time that was consumed during the execution of this scenario, we subtracted the estimated data communications time from the total execution time leaving only the time that was consumed by the remote processors while sorting their designated sub-arrays. This estimate was based on the average times of the different experiments that were performed on IMAN1. Figure-6 below presents the estimated results of these scenarios and compares it to the previous scenarios which were run on the single machine. Table-7 in Appendix [B] presents the timings details for this figure.

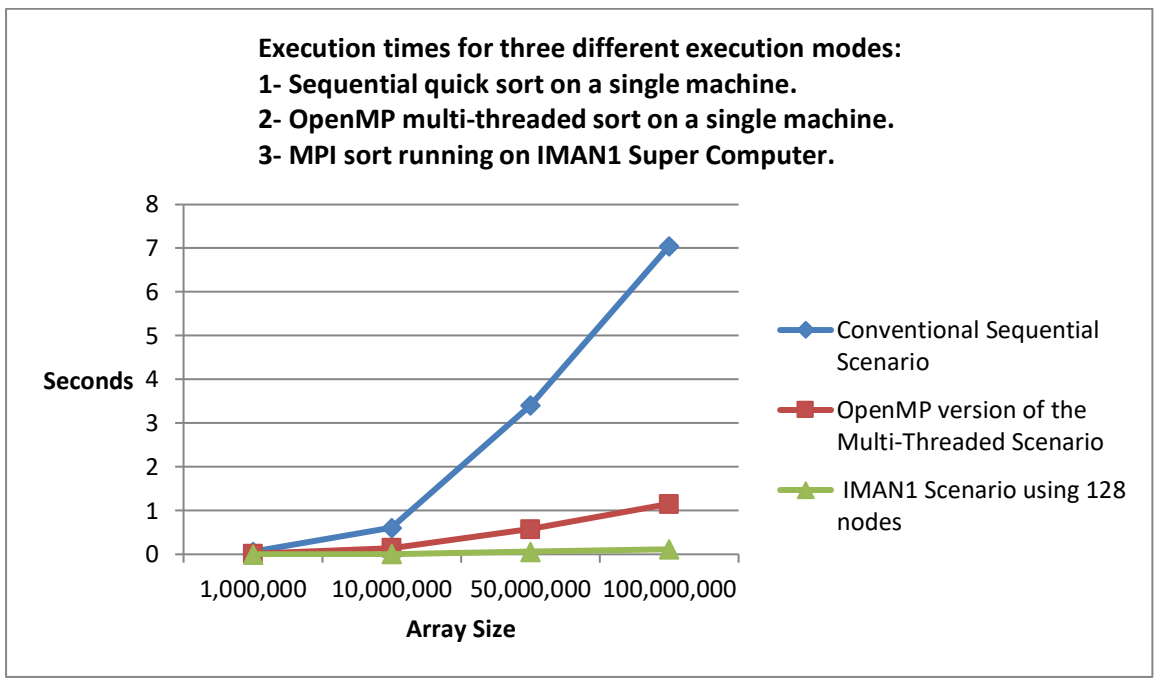

*Figure 6 – Execution times for three different scenarios while sorting arrays of different sizes, the conventional sequential quick sort, OpenMP multi-threading experiments using (8) threads of execution and the IMAN1 super computer MPI experiments using (128) nodes.*

In the following section of this paper, a discussion and analysis of the findings and results is presented to summarize the main outcomes of the experiments performed in this research.

**Analysis of Results**

The experiments performed by this research demonstrated that sequential sorting algorithms have inherent limitations especially when sorting large size arrays. Nevertheless, it was also demonstrated that the performance of these algorithms is significantly enhanced when using parallelism. Parallel computing can speed up sequential processing to different extents that are **< P** (number of processors) in general. The value of speed increase depends on various factors that include the used algorithm, the parallel architecture and the data set size.

The experiments demonstrated that parallel processing can increase execution speeds and to different degrees depending on the initial array size.

For example, under the (8) parallel threads and the (100.000.000) elements scenario, we compute *Speed Up* as:

*Speed Up = Serial Run Time / Parallel Run Time*

$$S = 15.303 / 7.161$$

$$S = 2.13 \text{ times faster}$$

While under the (8) parallel threads and the (400.000.000) elements

$$S = 30.576 / 4.883$$

$$S = 6.2 \text{ times faster}$$

The efficiency of the algorithm under the (100.000.000) elements array can be measured as:

*Efficiency = Speed / Number of Processors*

$$= 2.13 / 8$$

$$= .26 \%$$

While the efficiency of the algorithm under the (400.000.000) elements array was:

*Efficiency = Speed / Number of Processors*

$$= 6.2 / 8$$

$$= .775 \%$$

Knowing that in an ideal parallel system the *Efficiency* of the algorithm cannot exceed (1) which occurs when **Speed = Number of Processors**, the (.775%) ratio is considerably high and acceptable.

On the other hand, for the IMAN1 experimentations, abnormal results were noticed which was discussed in a previous section relating the reason mainly to the limited resources of the master node (256 MB RAM). Therefore, if we neglect the data preparation time and the data communications time at the master node, we can estimate the *Speed* and the *Efficiency* of the parallel algorithm for sorting (100.000.000) integers on a (128) nodes system following:

*Speed Up = Serial Run Time / Parallel Run Time*

$$= 7.35232 / 0.1184$$

$$= 64.04$$

*Efficiency = Speed / Number of Processors*

$$= 64 / 128$$

$$= 50\ \%$$

But if we considered data preparation and data communication times, the *Speed* and *Efficiency* would degrade noticeably.

In conclusion, it is known that a parallel algorithm is as fast as the slowest processor in its processors chain, and each parallel algorithm has inherent sequential parts that have to be executed sequentially even when distributed over numerous remote nodes. Therefore, parallel algorithms are not expected to expose linear performance increase when adding more computing resources, there will be a threshold where the communications costs will overcome the benefits of introducing more remote processing powers. In such scenarios, the observer will start to witness performance degrades after stepping beyond that threshold. This phenomena wasn't clearly visible in our experiments for different reason most importantly because of the nature of the used algorithm which performed all the initialization tasks at a central node before dispatching sub-arrays to remote processors. Also, this phenomenon was witnessed during the single machine scenarios because most of the communications were performed on the same machine using its high end internal memory for the communication between threads which resembled a local processing scheme much faster than any network enabled distributed processing. Nevertheless, it is expected that such performance degrade will be witnessed in the presented sorting algorithm especially in the super computer scenario, this degrade will start to appear once the algorithm crosses a threshold were the communications costs exceeds the parallelism benefits. Disclosing such a scenario would require more experiments on the super computer environment but only after fixing the limited capabilities issue of the master node.

Finally, it should be stated that the challenge will always exist to develop better parallel algorithms aiming to squeeze more computation powers from a better designed parallel architectures.

Source code: https://github.com/msfasha/Algorithms

**Appendix A, Execution Result sets**

| CPU Time in Seconds | | |
|---|---|---|
| Array Size | Sorted Arrays | Randomly Generated Arrays |
| 1,000,000 | 0.063 | 0.156 |
| 10,000,000 | 0.64 | 1.451 |
| 50,000,000 | 3.417 | 7.457 |
| 100,000,000 | 7.161 | 15.303 |
| 400,000,000 | 30.576 | 63.976 |

*Table 1 - The actual execution times gathered after running quick sort algorithm on a single machine using the conventional sequential version of quick sort. The table presents the results after running the algorithm over integer arrays of different ones.*

| Number of Iterations | | |
|---|---|---|
| Array Size | Sorted Arrays | Randomly Generated Arrays |
| 1,000,000 | 16,951,464 | 12,947,438 |
| 10,000,000 | 203,222,832 | 135,719,231 |
| 50,000,000 | 1,132,891,188 | 772,800,678 |
| 100,000,000 | 2,365,782,326 | 1,379,111,260 |
| 400,000,000 | 10,263,129,146 | 5,053,369,409 |

*Table 2 - The actual number of iterations gathered after running quick sort algorithm on a single machine using the conventional sequential version of quick sort. The table presents the number of iterations for the already-sorted arrays and for the randomly generated ones.*

| Number of Recursion Calls | | |
|---|---|---|
| Array Size | Sorted Arrays | Randomly Generated Arrays |
| 1,000,000 | 524,287 | 745,549 |
| 10,000,000 | 5,805,697 | 7,250,590 |
| 50,000,000 | 33,222,784 | 35,514,935 |
| 100,000,000 | 66,445,568 | 71,039,505 |
| 400,000,000 | 265,782,272 | 284,017,639 |

*Table 3 - The actual number of recursion calls gathered after running quick sort algorithm on a single machine using the conventional sequential version of quick. The table presents the number of recursion calls for the already-sorted arrays and for the randomly generated ones.*

| Number of Swap Operations | | |
|---|---|---|
| Number of Elements | Best Case Scenario | Random Case Scenario |
| 1,000,000 | 524,287 | 5,733,241 |
| 10,000,000 | 5,805,696 | 73,458,318 |
| 50,000,000 | 33,222,784 | 424,691,999 |
| 100,000,000 | 66,445,568 | 893,541,510 |
| 400,000,000 | 265,782,272 | 3,992,740,172 |

*Table 4 - The actual number of swap operations that was gathered after running quick sort algorithm on a single machine using the conventional sequential version of quick. The table presents the number of swap operations that was performed for the already-sorted arrays and for the randomly generated one.*

| Execution Times in Seconds for the Single Machine Three Scenarios | | | |
|---|---|---|---|
| Array Size | Sorted Arrays | Conventional multi-threading – 8 threads Sorted-Arrays | OpenMP multi-threading 8 threads Sorted-Arrays |
| 1,000,000 | 0.062 | 0.047 | 0.015 |
| 10,000,000 | 0.608 | 0.546 | 0.141 |
| 50,000,000 | 3.401 | 3.011 | 0.577 |
| 100,000,000 | 7.051 | 6.272 | 1.154 |
| 400,000,000 | 30.576 | 27.191 | 4.883 |

*Table 5 - The actual execution times gathered after running quick sort algorithm on a single machine under the 1-Conventional sequential version of quick sort, 2- The Convention multi-threaded version of the algorithm and under the 3-OpenMP version of the algorithm. The table presents the results after running these scenarios over integer arrays of different ones.*

| Execution Times in Seconds for Super Computer Scenarios | | | |
|---|---|---|---|
| Elements | Nodes | Preparation Time | Sending Time |
| 100000 | 15 | 0.120202 | 0.023271 |
| 1000000 | 15 | 0.869952 | 0.12327 |
| 1000000 | 31 | 0.819966 | 0.160007 |
| 1000000 | 63 | 0.823658 | 0.205206 |
| 5000000 | 7 | 4.188016 | 0.310268 |
| 5000000 | 15 | 4.075293 | 0.367887 |
| 5000000 | 31 | 4.067523 | 0.492961 |
| 5000000 | 63 | 4.096764 | 0.593166 |
| 5000000 | 127 | 4.155767 | 0.623767 |
| 10000000 | 15 | 12.172133 | 1.578022 |
| 10000000 | 31 | 9.800075 | 1.472349 |
| 50000000 | 7 | 121.6755 | 47.852055 |
| 50000000 | 31 | 220.894543 | 86.968016 |

*Table 6 - The actual execution times gathered after running quick sort algorithm on IMAN1 supercomputer. The table presents the time required to prepare and dispatch sub-arrays to the remote processors. This information presents these times for different scenarios related to different number of array sizes and different number of processors.*

| Execution Time in Seconds for three scenarios | | | |
|---|---|---|---|
| Array Size | Conventional Sequential Scenario | OpenMP version of the Multi-Threaded Scenario | Estimated IMAN1 Scenario using 128 nodes |
| 1,000,000 | 0.062 | 0.015 | 0.000963 |
| 10,000,000 | 0.608 | 0.141 | 0.007175 |
| 50,000,000 | 3.401 | 0.577 | 0.0574 |
| 100,000,000 | 7.051 | 1.154 | 0.1148 |
| 400,000,000 | 30.576 | 4.883 | 0.4592 |

*Table 7 - The actual execution times during the conventional sequential sort, OpenMP multi-threading and finally an estimated version for the IMAN1 super computer experiments.*

**Appendix B, Explanatory Figures**

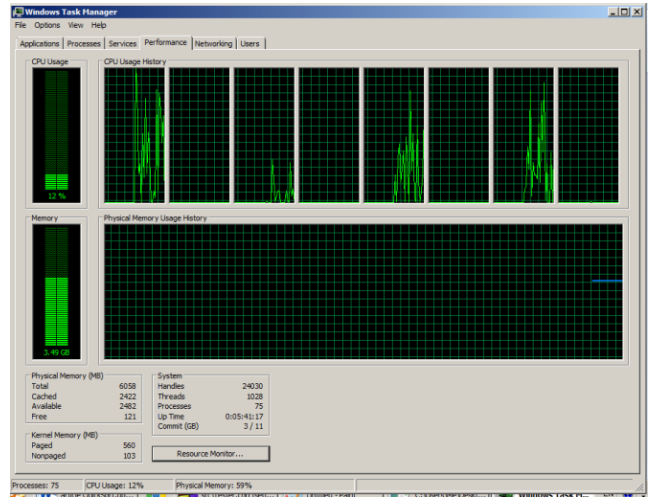

*Figure 1 – CPU utilization when running the conventional multi-threading mode over 8 threads of execution. The figure demonstrates that the CPU utilization was around 12% which reveals that a minimal part of the machine's processing powers were utilized.*

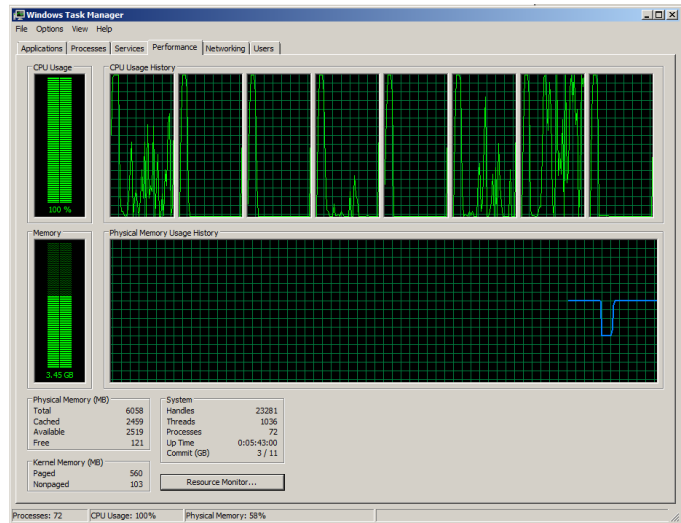

*Figure 2 – CPU utilization when running the OpenMP version of the multi-threading mode over 8 threads of execution. The figure demonstrates that the CPU utilization reached the maximum processing powers of the machine (100%).*